\documentclass[prc,aps,floatfix,groupedaddress,amsmath,amssymb]{revtex4}
\usepackage{graphicx}
\usepackage{epsfig}
\usepackage{dcolumn}% Align table columns on decimal point

\def\beq{\begin{equation}}
\def\eeq{\end{equation}}

\newcommand{\kv}{\mathbf{k}}

\newcommand{\kd}{{\rm k}}

\newcommand{\be}{\begin{eqnarray}}
\newcommand{\ee}{\end{eqnarray}}

\newcommand{\een}{\nonumber\end{eqnarray}}

\begin{document}

\title{Coupling between superfluid neutrons and superfluid protons in the elementary excitations of neutron star matter.}

\author{Marcello Baldo$^1$ and Camille Ducoin$^2$}
\affiliation{
$^1$ Dipartimento di Fisica, Universit\`a di Catania,
and INFN, Sezione di Catania,
Via Sofia 64, I-95123, Catania, Italy \\
$^2$ IPNL, CNRS/IN2P3, Universit\'e de Lyon/UCBL, F69622 Villeurbanne Cedex, France}

\begin{abstract}
Several phenomena occurring in neutron stars are affected by the elementary excitations that characterize the stellar matter. In
particular, low-energy excitations can play a major
role in the emission and propagation of neutrinos, neutron star cooling and transport processes. In this paper, we consider the elementary modes in the star region where both proton and neutron components are superfluid.
 We study the overall spectral functions of protons, neutrons and electrons on the basis of the Coulomb and nuclear interactions.
 This study is performed in the framework of the Random Phase Approximation, generalized to superfluid systems. The formalism we use ensures that the Generalized Ward's Identities are satisfied. We focus on the coupling between neutrons and protons. On one hand this coupling results in collective modes that involve simultaneously neutrons and protons, on the other hand it produces a damping of the excitations. Both effects are especially visible in the spectral functions of the different components of the matter.
At high density while the neutrons and protons tend to develop independent excitations, as indicated by the spectral functions,
the neutron-proton coupling still produces a strong damping of the modes.

%Despite their relative small fraction, the protons turn out to modify the neutron spectral function as a consequence of the nuclear %neutron-proton interaction. This effect is particularly evident at the lower density, just below the crust for a density close to %the saturation value, while at increasing density the neutrons and the protons are mainly decoupled. The proton spectral function %is characterized by a pseudo-Goldstone mode below $ 2\Delta $, twice the pairing gap, and a pair-breaking mode above  $ 2\Delta $. %The latter merges in the sound mode of the normal phase at higher momenta. The neutron spectral function develops a collective %sound mode only at the higher density. The electrons have a strong screening effect on the proton-proton interaction at the lower %momenta, and decouple from the protons at higher momenta.    
\end{abstract}

\maketitle

%%%%%%%%%%%%%%%%%%%%%%%%%%%%%%%%%%%%%%%%%%%%%%%%%%%%%%%%%%%%%%%%%%%%%%%%%%%%%%
%%%%%%%%%%%%%%%%%%%%%%%%%%%%%%%%%%%%%%%%%%%%%%%%%%%%%%%%%%%%%%%%%%%%%%%%%%%%%%
\section{Introduction}
%%%%%%%%%%%%%%%%%%%%%%%%%%%%%%%%%%%%%%%%%%%%%%%%%%%%%%%%%%%%%%%%%%%%%%%%%%%%%%
%%%%%%%%%%%%%%%%%%%%%%%%%%%%%%%%%%%%%%%%%%%%%%%%%%%%%%%%%%%%%%%%%%%%%%%%%%%%%%

In neutron stars, many phenomena and processes that occur in the outer and inner core are affected by the presence of matter elementary excitations. In particular these excitations  have a relevant role in neutrino emission and propagation, specific heath and transport phenomena, which are all involved in the short and long time evolution of the star. 
At not too high density it is expected that the main components of the matter are neutrons, protons, electrons 
and muons~\cite{shap}, and then
the spectral properties of these excitations can have a complex structure.
Collective modes in asymmetric nuclear matter have been studied previously,
e.g. in Refs.~\cite{Haensel-NPA301, Matera-PRC49, Greco-PRC67}.
In the astrophysical context, a study of the collective excitations in normal neutron star matter
within the relativistic mean field method has been presented in Ref.~\cite{Providencia-PRC74}.
The overall spectral functions of the different components in normal neutron star matter
have been calculated in Ref.~\cite{paper1,paper2} on the basis of non-relativistic Random Phase Approximation (RPA)
for the nucleonic components and relativistic RPA for the leptonic components.
Different models for the nuclear effective interaction were considered
and a detailed comparison was done between some Skyrme forces and a microscopically derived interaction.
 More recently the relativistic RPA
has been employed also for the nucleon components \cite{Stet}.
The works of ref. \cite{paper1,paper2} were extended to superfluid proton matter in ref. \cite{paper3}, but neglecting the proton-neutron coupling.
The elementary excitations in superfluid neutron star matter have been analyzed by several authors
~\cite{Reddy,Kundu,Leinson1,Armen,Leinson2,Vosk}.
In ref. \cite{paper3} it was shown that the proton superfluid matter, due to the electron screening of the proton-proton Coulomb interaction, still presents a pseudo-Goldstone mode at low momentum. Its presence can have a strong influence e.g. on neutrino emission~\cite{Yako,Reddy,Kundu,Leinson1,Armen,Leinson2,Vosk} or mean free path.
More recently in refs. \cite{Urban1,Thesis} the elementary excitations of superfluid neutron matter in the crust region were studied, and with the possible inclusion of the coupling with the nuclear lattice in refs. \cite{Chamel,Urban2,Thesis,Koby} .    
\par
In this paper we focus on the region of homogeneous core matter where both neutron and proton superfluidity can occur. The region of the neutron stars where this indeed happens is not well determined and it could even be that neutron and proton superfluidity
never coexist. Therefore we explore different densities and different pairing gaps for neutrons and protons in order to
figure out the possible scenarios that can be expected in the neutron star matter. We study the effect of the neutron-proton
interaction, under which conditions neutrons and protons are coupled and to what extent they can be simultaneously excited. An extensive study of the elementary excitations in presence of both proton and neutron superfluidity has been presented in ref. \cite{Koby1}, where the hydrodynamics formalism was used with the inclusion of proton-neutron coupling.
As in ref. \cite{paper3} we introduce the general theoretical scheme within the generalized RPA approximation, which is known to be a conserving approximations~\cite{BaymKad,Baym}, i.e. current is conserved locally and the related Generalized Ward's Identities\cite{Schriefferb} are fulfilled. The formalism can be derived by different methods, e.g. by the equation of motion technique
\cite{Gusakov} or the functional derivative scheme \cite{paper1,paper2,paper3}. In any case the basic equations correspond to the generalized RPA. 
One of the main goal of our work is the study of the mutual influence of protons and neutrons on the overall spectral functions. 
 This work extends the study of refs. \cite{paper4} by including also the neutron superfluidity. 

\par
The plan of the paper is as follows. In Sec. \ref{sec:form} the formalism for the response function in the generalized RPA scheme is briefly sketched. In particular it is discussed the method to estimate microscopically the effective nucleon-nucleon interaction. In Sec. \ref{sec:res} the results are presented for the spectral function taking the neutron and proton pairing gaps as parameters. The role of the neutron-proton interaction is discussed in detail. In Sec. \ref{sec:conc} we summarize and draw the conclusions. Finally in the Appendix additional details of the calculations are given. 
\section{Formalism. \label{sec:form}}
For completeness, the generalized Random Phase Approximation is here briefly discussed. In multi-component fermion systems the equations for the generalized response functions $ \Pi $ can be written schematically \cite{paper3,paper4,Schriefferb}
\beq
\Pi_{ik}(t,t') \,=\, \Pi_{ik}^{0}(t,t') \,+\, \sum_{jl} \,  \Pi_{ij}^{0}(t,\overline{t_1}) v_{j,l} \Pi_{lk}(\overline{t_1},t') 
\label{eq:RPA}\eeq
\par\noindent
where $ \, i,j, \cdots \,$ label the different components and the corresponding degrees of freedom, $\,  v_{j,l}\, $ is the effective interaction between them and $\, \Pi^{0}\, $ is the free response function. The time variable with an overline is integrated.
In neutron star matter one has neutron, proton and electron components (neglecting muons),
with the possibility of both particle-hole and pair excitations in the nucleon channels. Since we assume the presence of both proton and neutron pairing, in terms of creation and annihilation operators the indexes $\, i,j, \cdots $ include the following configurations
\beq
\begin{array}{ll}
\ &\ a^\dag(p)\,a(p) | \Psi_0>\,\,\ ,\ \,\, a^\dag(p)\,a^\dag(p) | \Psi_0>\,\,\ ,\ \,\, a(p)\,a(p) | \Psi_0> \\
\ &\ \\
\ &\ a^\dag(n)\,a(n) | \Psi_0>\,\,\ ,\ \,\, a^\dag(n)\,a^\dag(n) | \Psi_0>\,\,\ ,\ \,\, a(n)\,a(n) | \Psi_0> \\ 
\ &\ \\
\ &\ a^\dag(e)\,a(e) | \Psi_0>  
\end{array}
\label{eq:conf}\eeq
\par\noindent
where the labels $\, n, p, e\, $ indicate neutrons, protons and electron respectively, and $ |\Psi_0> $ is the correlated ground state. If we call $ A_i | \Psi_0>$ the generic configuration, the response functions can be written
\beq
 \Pi_{ik}(t,t') \,=\, - < \Psi_0 | T\{A_i^\dag(t) A_k^{\phantom{\dag}}(t')\} | \Psi_0 > 
\label{eq:Pi}\eeq 
\par\noindent
where $ T $ is the usual fermion time ordering operator.
The configurations (\ref{eq:conf}) correspond in fact to both density and pairing excitations. In agreement with (\ref{eq:conf}), Eq. (\ref{eq:RPA}) forms in general a $ 7\times 7 $ system of coupled equations. However it turns out \cite{paper3,paper4,Schrieffer} that two equations can be decoupled to a good approximation by taking suitable linear combinations of the pairing additional mode $ a^\dag\, a^\dag $ and pairing removal mode $ a\, a $ for both neutrons and protons. In this way the system reduces to  $ 5\times 5 \, $ coupled equations. Details on the equations and their explicit analytic form are given in Appendix A. The system has to be solved for the response functions $\, \Pi_{ik}\, $, all of which can be obtained by selecting the inhomogeneous term in Eq. (\ref{eq:RPA}). More precisely one has to select a given configuration indicated by the right index $ k $ and solve the system for each choice of $ k $. In this way one gets all the diagonal and non-diagonal elements of $\, \Pi_{ik}\, $.  
\par
One has to notice that the generalized RPA equations are valid in the collisionless regime, so that the only damping of the modes is the Landau damping, which is very effective above a certain momentum threshold. In particular dissipation due to electron-electron collisions is neglected. This is justified if the electron mean free path is much larger than the typical wavelength of the mode. Under the physical conditions of neutron star matter, i.e. low temperature and density of the order of the saturation one, the electron mean free path was estimated in ref. \cite{Shtern}, where it was shown that the collisions are dominated by the exchange of transverse plasmon modes and the mean free path extends to a macroscopic size, of the order of 10$^{-3}$ cm.
This also indicates that the collision time is much longer than the characteristic period of the modes, and that the electron collisions are relevant only for macroscopic motion like viscous flow. We therefore neglect in the sequel electron dissipation, and include only the Landau damping.           
\par 
For simplicity, for the single particle energy spectrum we are going to use the bare mass. The introduction of effective masses is trivial and we think that it is not going to change qualitatively the overall pattern of the results, but of course for quantitative results the effective mass is mandatory. Then the main input needed in Eq. (\ref{eq:RPA}) is the effective interaction $ v_{ij} $. The pairing interaction strength $ U $ is very sensitive to many-body effects \cite{ppair} and it is quite challenging to estimate its size. We prefer to use the pairing gap as a parameter to be explored and fix the pairing interaction consistently with the gap equation  ( $ U > 0 $ )
\be
\Delta &=& U\int{\frac{d^3 \kv}{(2\pi)^3}\frac{\Delta}{2E_{\kd}}}
= U\int{\frac{d^3 \kv}{(2\pi)^3} u_{\kd}v_{\kd}}
\label{eq:gap}\ee
\noindent both for neutrons and protons. The quasi-particle energy $ E $  and coherence factors  $ u, v $ have the standard form
\be
\begin{array}{ll}
 E_{\kd} &\,=\, \sqrt{(\epsilon_{\kd} - \mu)^2 + \Delta^2} \\
\ &\ \\
 v_{\kd}^2 &\,=\, \frac{1}{2} \big( 1 \,-\, \frac{\epsilon_{\kd} - \mu}{E_{\kd}} \big) 
 \ \ \ \ \ , \ \ \ \ \   u_{\kd}^2 \,=\, \frac{1}{2} \big( 1 \,+\, \frac{\epsilon_{\kd} - \mu}{E_{\kd}} \big) \\  
\end{array}
\label{eq:stand}\ee
\noindent where $ \epsilon_{\kd} $ is the kinetic energy and $ \mu $ the chemical potential. The necessary cut-off in momentum for the integration of Eq. \ref{eq:gap} has been fixed to $\sqrt{2} k_F $, where $k_F$ is the proper Fermi momentum, as in our previous works \cite{paper2,paper3}. In the calculations the value of the pairing gaps $ \Delta_{n,p} $ are fixed at given values and the effective interaction strengths $ U_{n,p} $ are extracted from the corresponding gap equation (\ref{eq:gap}). Then the pairing interaction $ -U $ is inserted in the RPA equations (\ref{eq:RPA}). \par 
For the various particle-hole interactions we focus on the density response function, i.e. the vector channel, and then we follow the Landau monopolar approximation. The corresponding strengths can be estimated on the basis of a realistic Skyrme interaction.
It is also possible to consider the microscopic many-body Equation of State (EOS) as an Energy Density Functional. In this approach for  Brueckner-Hartree-Fock (BHF) calculations the interaction strength can be obtained from the derivative with respect to the density $ \rho_j$ of the BHF potential energy $ V_i $, with $ i, j $ running on the proton and neutron components
%In the Landau monopolar approximation, we can identify the residual interaction
%with the derivative of the single-particle potential at constant momentum, taken at Fermi level:
%%%
\beq
\label{eq:vres-derU}
v_{ij}=\left(\frac{\delta V_i}{\delta\rho_j}(k_{{\rm F}i},\rho_n,\rho_p)\right)_{k,\rho_i={\rm cst}}\;.
\eeq
%%%    
\noindent Notice that in Eq. (\ref{eq:vres-derU}) 
%the Fermi momenta $ k_{{\rm F}i}$ 
%must be kept fixed in performing the derivative.
%How to do that in practice from the numerical calculations is explained in ref. \cite{paper1}. 
%%%%%%%%%%%%%
%Camille
the Fermi momenta $ k_{{\rm F}i}$ are kept fixed in performing the derivative,
in order to separate the kinetic contribution associated with effective-mass effects. 
More details on that point are given in ref. \cite{paper1}. 
%%%%%%%%%%%%%
In the case of Skyrme forces the possible dependence on momenta is analytic and the procedure is trivial. In any case no strong pairing effects are considered,
e.g. the effective interactions are assumed to be independent of the pairing gap and calculated for the normal system.
\section{Results.\label{sec:res}}
The calculations have been performed by including nuclear pairing, Coulomb, and density-density nuclear interactions but we first
summarize some of the standard results when only some of these interactions are included. They are well known in the literature, but they can be useful in guiding the interpretation of the general results. 
\par
It is well known that in a neutral superfluid, with only pairing interaction, there are two types of excitations. Below $ 2\Delta $
a sharp Goldstone mode is present. It is a consequence of the breaking of gauge invariance that occurs in the ground state of a superfluid system, and its energy is linear in momentum for small momenta, with a velocity equal to $ v_F/\sqrt{3} $, being $ v_F $ the Fermi velocity. Notice that a gauge transformation on the field operators is equivalent to a $ U(1) $ transformation on the order parameter \cite{Greiter}. This exitation is usually referred to as "superfluid phonon".
At increasing momentum the energy spectrum deviates from linearity and approaches $ 2\Delta $ for large momenta \cite{paper2}. Above $ 2\Delta $      
another excitation mode appears, usually indicated as " pair breaking "  mode, because it corresponds indeed to the breaking of a Cooper pair. It is strongly damped and it is reflected in a bump of the spectral function above $ 2\Delta $. The energy of this mode, after increasing with the momentum, bends down towards $ 2\Delta $ also. At even higher momentum the spectral function has no structure and any excitation is overdamped \cite{paper2}. \par 
If we consider only the proton component and introduce the Coulomb interaction, in principle the Goldstone mode should disappear,
and it should be substituted by a proton plasma mode which has a finite energy at zero momentum. However it has been shown in ref. \cite{paper2} that the electrons are fully screening the proton-proton Coulomb interaction, and a sound mode reappears below $ 2\Delta $. The (screened) Coulomb interaction however affects the sound velocity, which turns out to be about three times the Goldstone mode velocity. This mode can be considered a pseudo-Goldstone mode, since it is still below $ 2\Delta $ but with a modified velocity due to the interaction. Details can be found in ref. \cite{paper2}, where the structure of the proton spectral function is discussed in detail. 
%Here we notice only that the electron Thomas-Fermi screening length $v_F / \sqrt{3}\omega_p $, where $\omega_p$ is the electron plasma frequency and $v_F$ the Fermi velocity, at saturation density and with the calculated proton fraction is about 18 fm and about 11 fm at twice saturation density.
%%%%%%%%%%%
%Camille
Here we only comment on the scale of the electron Thomas-Fermi screening length. This length is given by $v_F / \sqrt{3}\omega_p $, where $\omega_p$ is the electron plasma frequency and $v_F$ the Fermi velocity:
with the calculated proton fraction, it is about 18 fm at saturation density and about 11 fm at twice saturation density. 
%%%%%%%%%%%
This is more than one order of magnitude larger than the average interparticle distance. However the Coulomb interaction is in any case screened, and this is enough to suppress the proton plasma mode. \par 
We now introduce the nuclear interaction, including the neutron pairing and the proton-neutron nuclear coupling. The proton fraction is taken from BHF calculations, which include three-body forces and correctly reproduce the phenomenological saturation point. The corresponding nuclear interaction strengths are calculated according to Eq. (\ref{eq:vres-derU}). The values of these physical parameters are reported in ref. \cite{paper1}. For simplicity we assume that the neutron pairing is in the $^{1}$S$_0$ channel.
The case of neutron pairing in the $^{3}$P$_2$ channel requires a separate treatment, due to the complexity of the possible neutron excitations.\par 
The spectral or strength function $ S(q,\omega) $, at given momentum $ q $ and energy $ \omega $, is directly related to the response function $ \Pi $
\beq
S_i(q,\omega) \,=\, -{\rm Im} \big( \Pi_{ii}(q,\omega) \big) 
\label{eq:strength}\eeq
\noindent where $ {\rm Im} $ indicates the imaginary part and the index $ i $ runs over the particle-hole configurations of neutrons, protons and electrons. It has to be stressed that the particle-hole configurations are coupled to the pairing channels, according to Eqs. (\ref{eq:RPA},\ref{eq:conf}), and indeed the non-diagonal matrix elements between particle-hole and pairing configurations of the response function are non-zero. The strength function is therefore dependent on the pairing gaps. 
\par 
We start our analysis in the case that the neutron gap is larger than the proton gap. As a representative matter density we
choose the symmetric matter saturation density $ \rho \,=\,$ 0.16 fm$^{-3}$, for which the proton fraction is 3.7\%.
This density is expected to correspond to the region just below the neutron star crust. The proton gap $\Delta_p$ is taken equal to 1 MeV and the neutron gap $\Delta_n$ equal to 1.5 MeV. These values do not correspond to any theoretical estimate, but they are just used to explore the qualitative scenario that can occur in case both neutrons and protons are in the superfluid phase. The gross features of the results should not depend on the particular values
of the gaps, provided they are not vanishing small.  
In Fig. 1a  are reported the strength functions of the three matter components at the momentum
$ q\, =\, $0.0125 fm$^{-1}$ and no neutron-proton interaction. The neutron and proton components are independent and one can recognize the pseudo-Goldstone (phonon) mode for the neutron component below 2$\Delta_n$ (black thick line). The position of the mode is represented by a sharp line, since it is actually a delta-function. Just above 2$\Delta_n$ one can see the pair-breaking mode, which has a wide distribution. Similarly for the proton components (red thin line) both modes are apparent. However in this case the pseudo-Goldstone has a large width. As discussed in ref. \cite{paper3,paper4} this is due to the Coulomb coupling of the protons with the electrons (green dashed line). In fact the electron strength function follows closely the proton strength function and the position of the mode is inside the electron particle-hole continuum where Landau damping is active. \par 
In Fig. 1b the same strength functions are reported when the neutron-proton coupling is switched on. The neutron-proton interaction
has different effects on the pseudo-Goldstone mode. In the energy region of this mode all three components are simultaneously excited, so the phonon is just a neutron-proton-electron excitation. The coupling produces a small shift of the neutron peak to a smaller energy and at the same time the whole mode acquires a substantial width. The width can be considered as induced by the width of the proton and electron components. The proton strength is strongly modified and partly shifted towards the neutron strength peak position, 
%%%%%%%%%%%
%Camille
forming a double peak.
The electron strength still follows closely the proton one. 
%%%%%%%%%%%
The pair-breaking modes look only marginally affected.
\par        
The presence of a width $W$ in the phonon strength function can have a drastic effect on the processes where they are involved.
In fact the phonon has a finite lifetime $\tau$ and it can propagate only for a finite distance $l$. An estimate of this distance can be obtained by multiplying the lifetime by the phonon velocity $v_G$
\beq
l \,=\, \tau v_G \,=\, \frac{\hbar}{W} v_G 
\eeq    
\noindent Assuming a width of a fraction of MeV and a phonon velocity of few times the neutron Fermi velocity one gets an order of magnitude estimate for $l$ of few hundreds fm. This is a microscopic distance and all phenomena involving macroscopic phonon mean free path are suppressed.\par
\begin{figure}
\vskip -8 cm
\includegraphics[bb= 320 0 470 790,angle=0,scale=0.55]{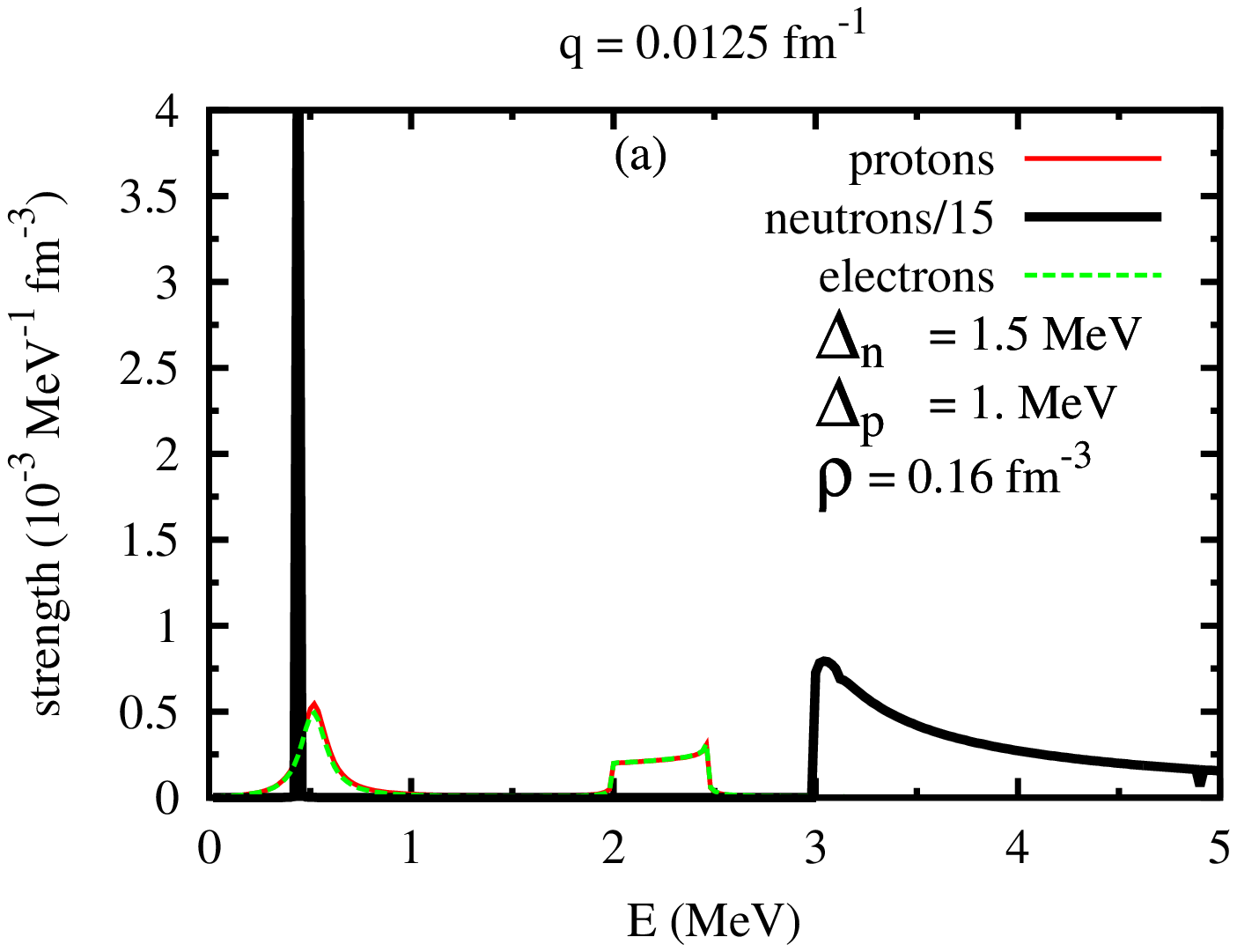}
\includegraphics[bb= 60 0 230 790,angle=0,scale=0.55]{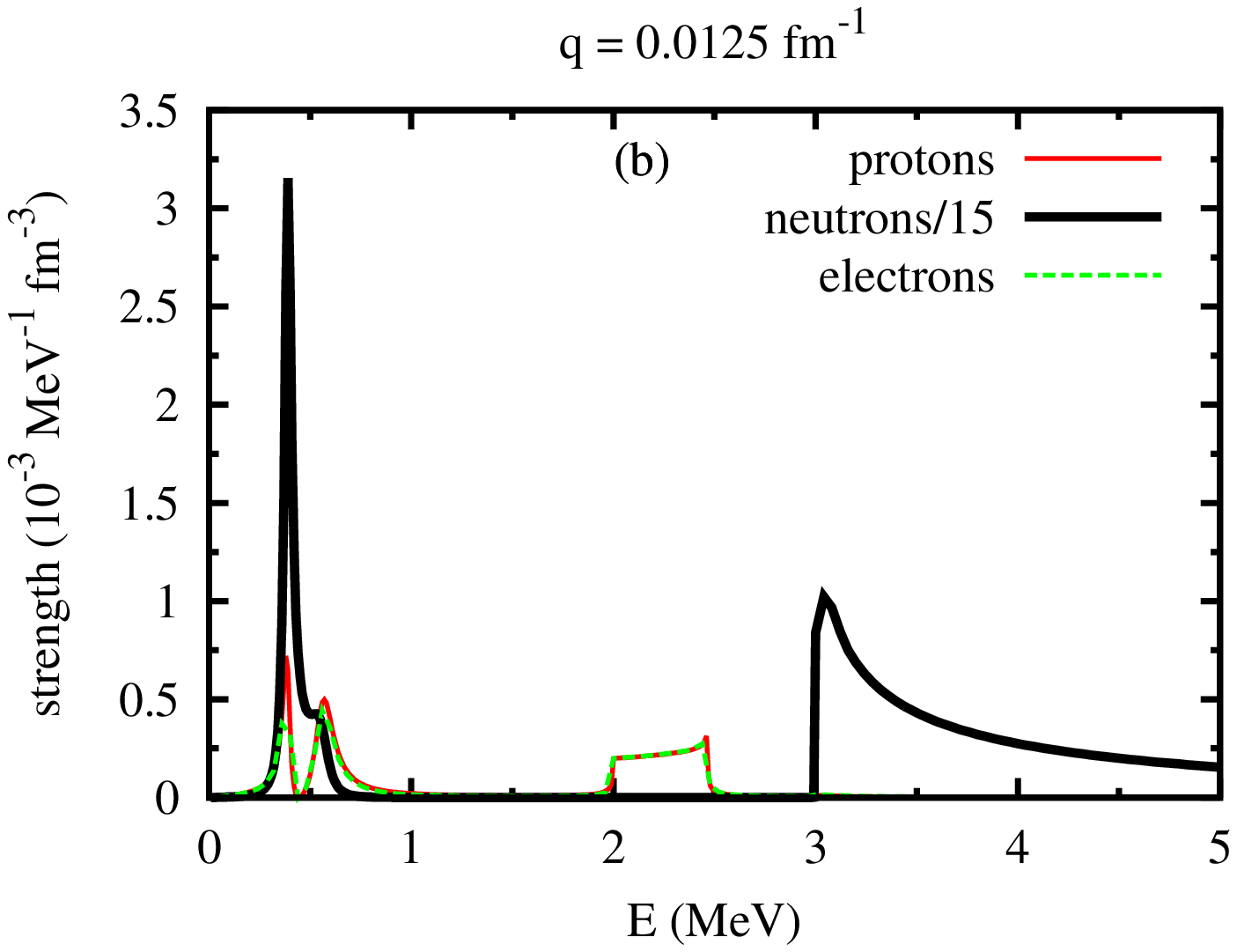}
\vskip -1 cm
\caption{(Color on line) Spectral functions of neutrons (black thick line), proton (red thin line) and electrons (green dashed line) calculated at the total baryon density $\rho = 0.16$ fm$^{-3}$. The neutron and proton pairing gaps are 1.5 MeV and 1 MeV, respectively. In panel (a) the neutron-proton interaction has been suppressed, while in panel b) it is included. For convenience the neutron strength function has been divided by 15.}
\label{fig:Fig1}
\end{figure}      
\par
\begin{figure}
\vskip -8 cm
\includegraphics[bb= 320 0 470 790,angle=0,scale=0.55]{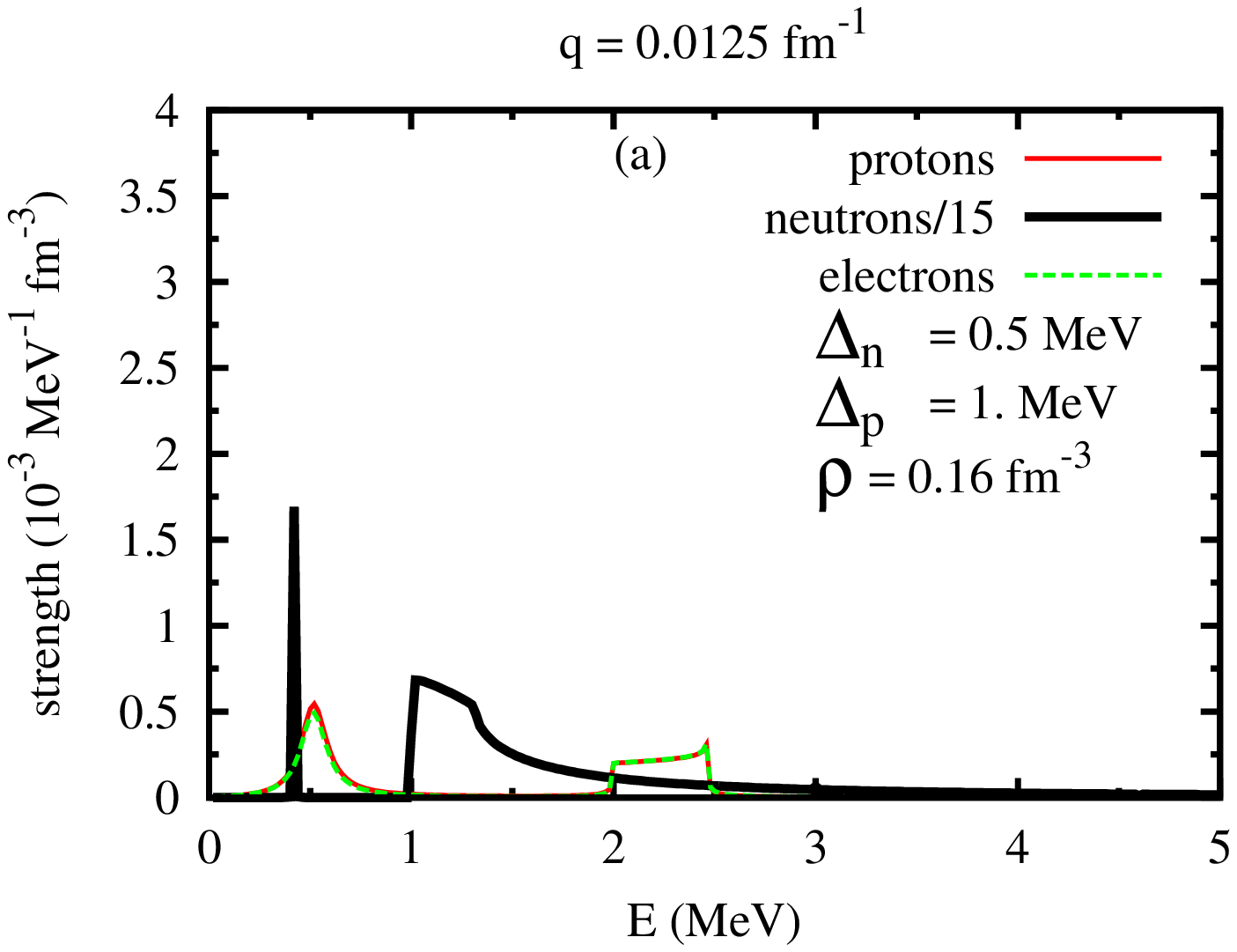}
\includegraphics[bb= 60 0 230 790,angle=0,scale=0.55]{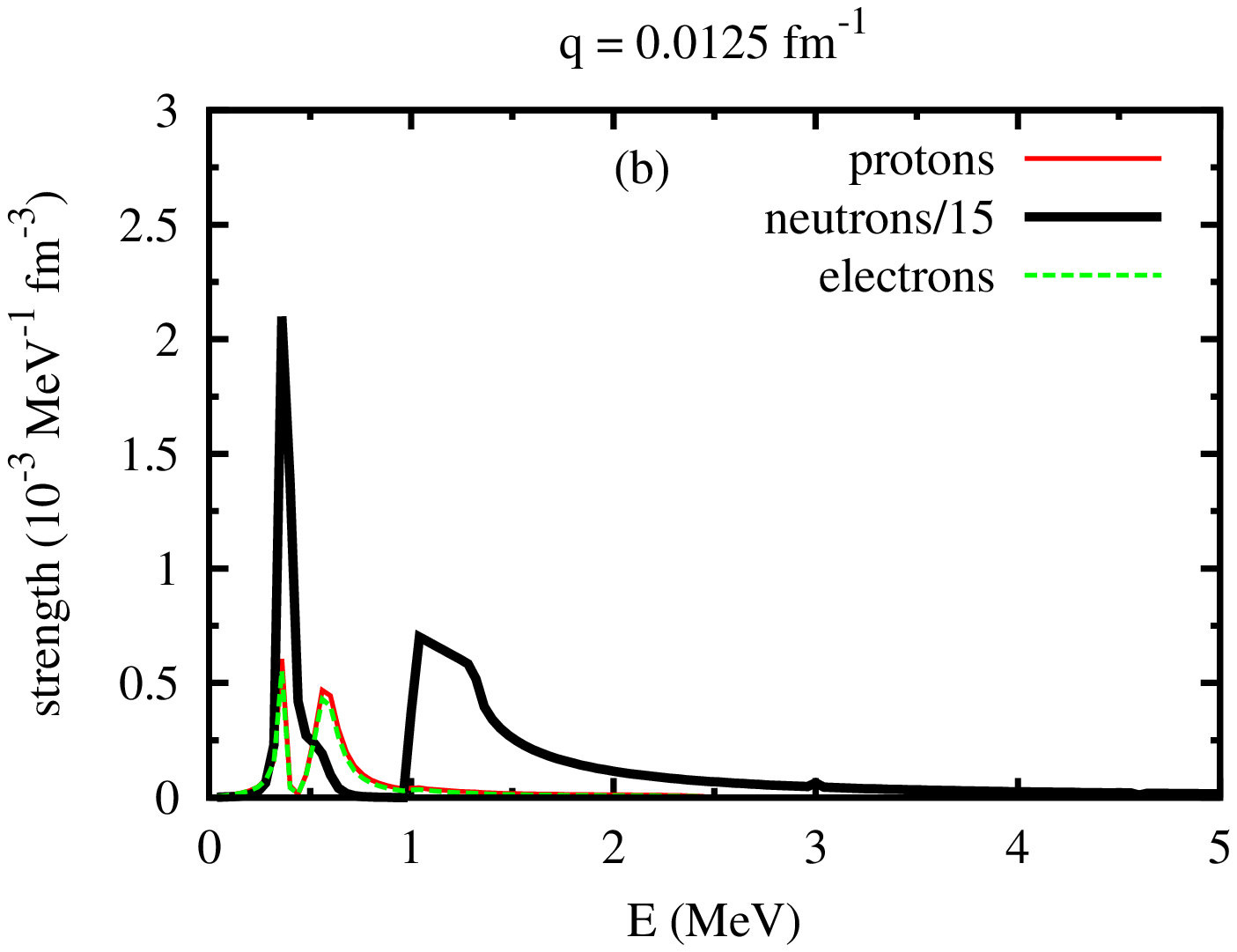}
\vskip -1 cm
\caption{(Color on line) Spectral functions of neutrons (black thick line), proton (red thin line) and electrons (green dashed line) calculated at the total baryon density $\rho = 0.16$ fm$^{-3}$. The neutron and proton pairing gaps are 0.5 MeV and 1 MeV, respectively. In panel (a) the neutron-proton interaction has been suppressed, while in panel b) it is included. For convenience the neutron strength function has been divided by 15.}
\label{fig:Fig2}
\end{figure}

 
In Fig. \ref{fig:Fig2} the neutron pairing gap has been reduced to 0.5 MeV, below the proton pairing gap of 1 MeV. A similar trend is observed. Also in this case the proton strength function is shifted to lower energy by the neutron-proton interaction, with a characteristic double peak structure. In addition the proton pair-breaking mode, which is apparent without neutron-proton coupling, is suppressed when the interaction is introduced. 
\par
\begin{figure}
\vskip -1 cm
\includegraphics[bb= 300 0 470 790,angle=0,scale=0.55]{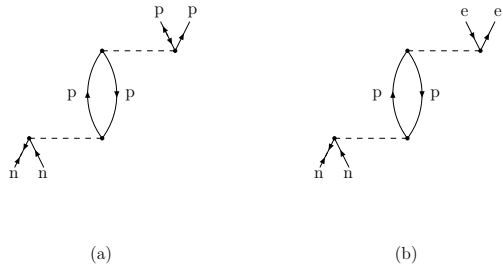}
\vskip -9.5 cm
\caption{Diagram a) illustrates the coupling between the neutron and proton pairing gap fluctuations. Diagram b) illustrates the coupling between the neutron pair fluctuations and the electron excitations. See the discussion in the text. The labels n,p,e stand for neutrons, protons and electrons, respectively.}
\label{fig:Fig3}
\end{figure}

The superfluid phonons of Figs. \ref{fig:Fig1}b and \ref{fig:Fig2}b correspond to the simultaneous fluctuation of neutron and proton pairing gaps. The reason of the coupling between the two fluctuations can be understood from the diagram of Fig. \ref{fig:Fig3}a, which is automatically included in the RPA equations. In this diagram a line with a double arrow indicates an anomalous propagator. One can see that the propagation of a neutron pair is coupled to a proton pair even if there is not a direct interaction between the two pairs, which is due to the anomalous neutron and proton propagators and the particle-hole neutron-proton interaction. Any fluctuation of the neutron pairing gap is therefore coupled to the fluctuations of the proton pairing gap, and 
the other way around.\par 
Similarly one can understand the reason of the phonon width by considering the diagram of \ref{fig:Fig3}b, also included in the RPA equations. One can see that the neutron pairs are coupled indirectly to the electron particle-hole excitations. Since the latter are inside the Landau damping region, the overall mode is damped through the neutron-proton interaction.
\par
\begin{figure}
\vskip -8 cm
\includegraphics[bb= 320 0 470 790,angle=0,scale=0.55]{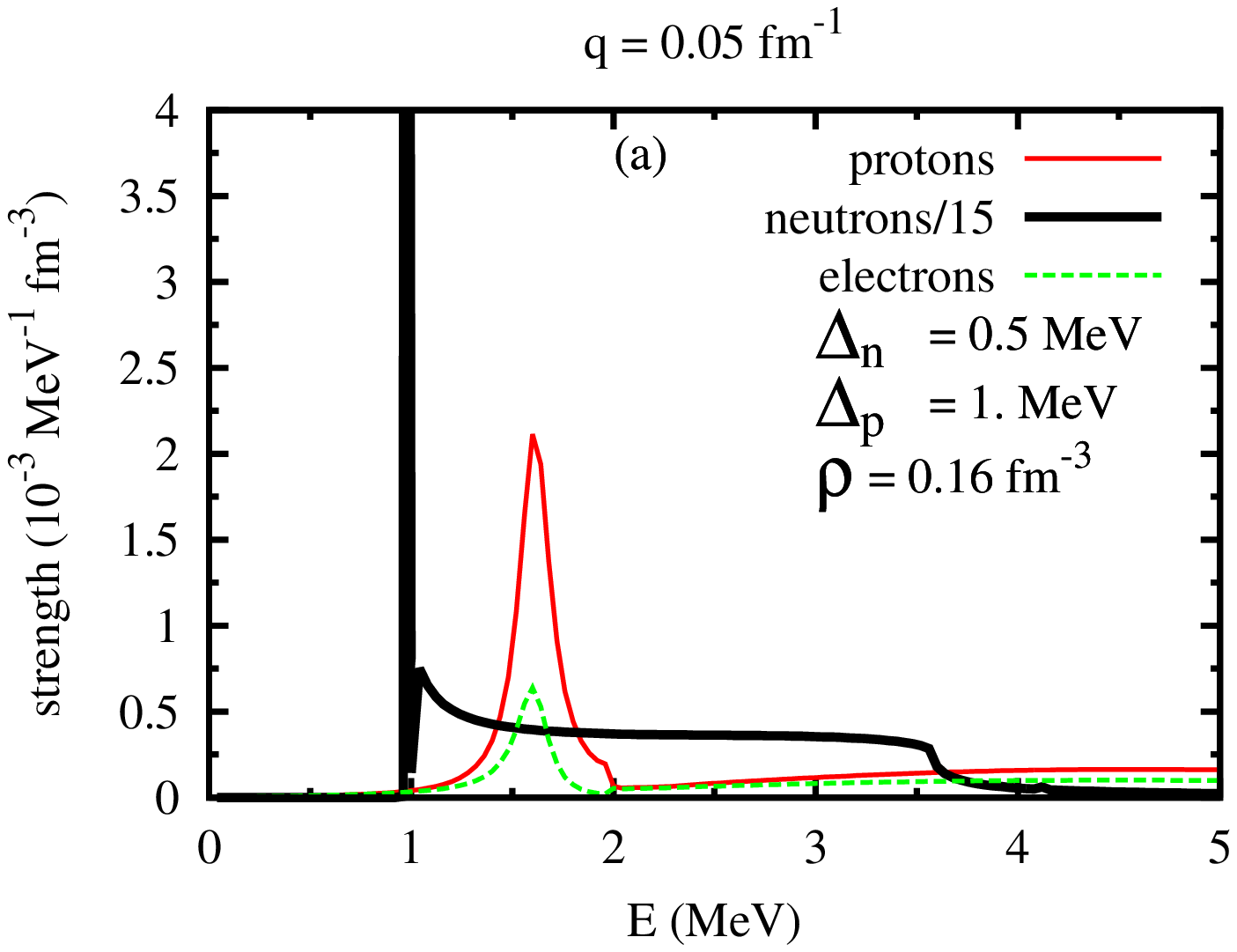}
\includegraphics[bb= 60 0 230 790,angle=0,scale=0.55]{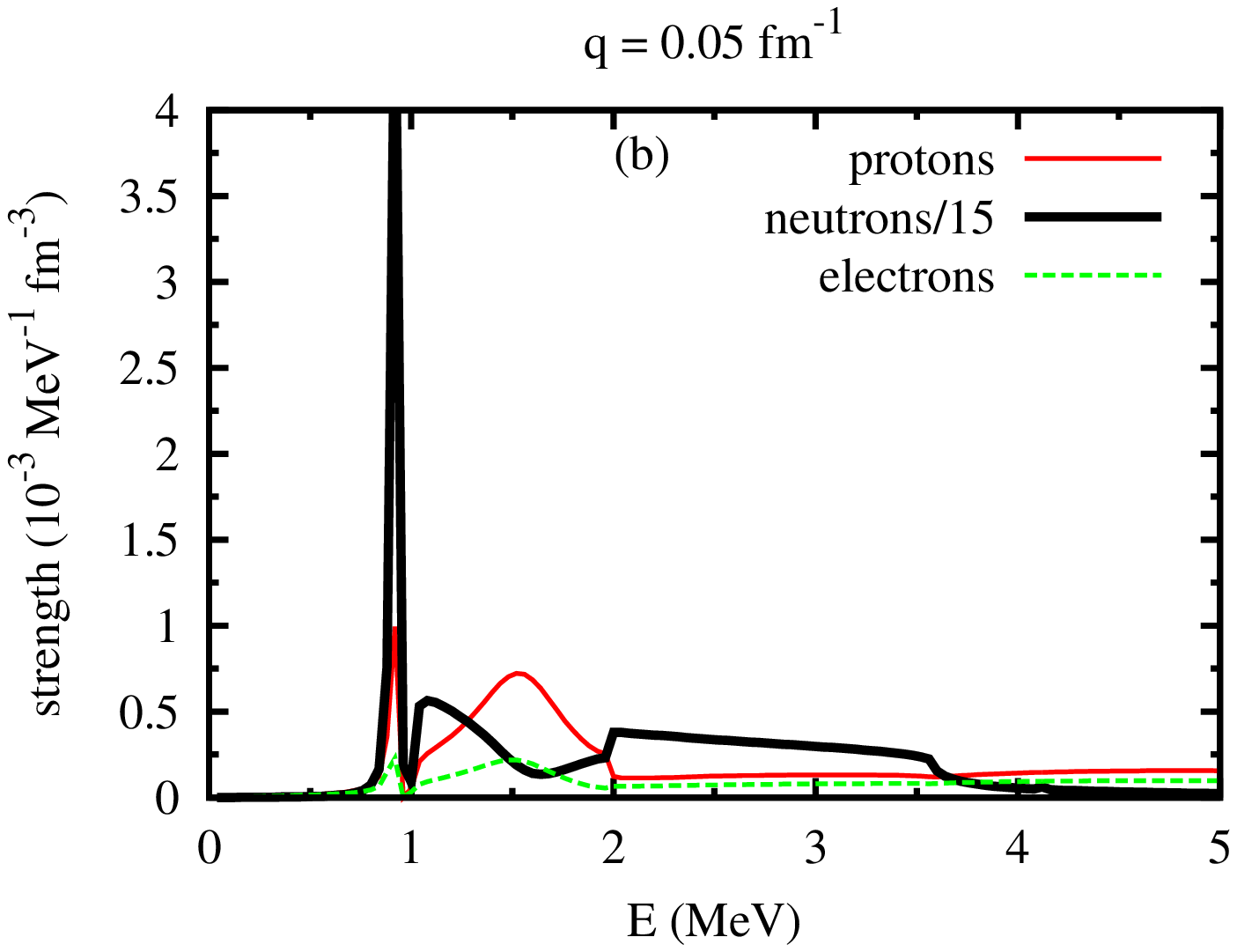}
\vskip -1 cm
\caption{(Color on line) Spectral functions of neutrons (black thick line), proton (red thin line) and electrons (green dashed line) calculated at the total baryon density $\rho = 0.16$ fm$^{-3}$ with the same pairing gaps as in Fig. \ref{fig:Fig2}, but at higher momentum. In panel (a) the neutron-proton interaction has been suppressed, while in panel b) it is included. For convenience the neutron strength function has been divided by 15.}
\label{fig:Fig4}
\end{figure}

 
As already mentioned, at higher momentum the energy of the phonon moves towards twice the pairing energy, where it joins with the pair-breaking mode. The spectral function changes accordingly. This can be seen in Fig. \ref{fig:Fig4}. In panel (a) are  reported the spectral functions at $ q \,=\, $0.05 fm$^{-1}$, with the same pairing gaps and at the same density, in the case of no neutron-proton interaction. One can see that now the phonon position, marked by a sharp line, is very close to 2$\Delta_n$. The independent proton phonon is still below 2$\Delta_p$, but as expected is substantially damped by the coupling with the electrons.
Notice that the electrons are now not able to follow closely the proton oscillations, due to the higher frequency of the mode, and the electron strength function is substantially smaller than the proton one. \par 
When the neutron-proton coupling is switched on, the phonon is only slightly shifted downward, it acquires a small proton component and it is substantially damped. Notice however that an appreciable proton strength function remains at the position of the previous proton phonon, but with a broad distribution. This indicates that the proton-neutron coupling is effective also in the case that the original uncoupled proton and neutron strength functions, panel 4(a), do not overlap appreciably. It is clear that the proton strength function is quite fragmented. The electron strength function is further reduced. In any case the strength function is dominated by the phonon peak.      
\par
\begin{figure}
\vskip -8 cm
\includegraphics[bb= 320 0 470 790,angle=0,scale=0.55]{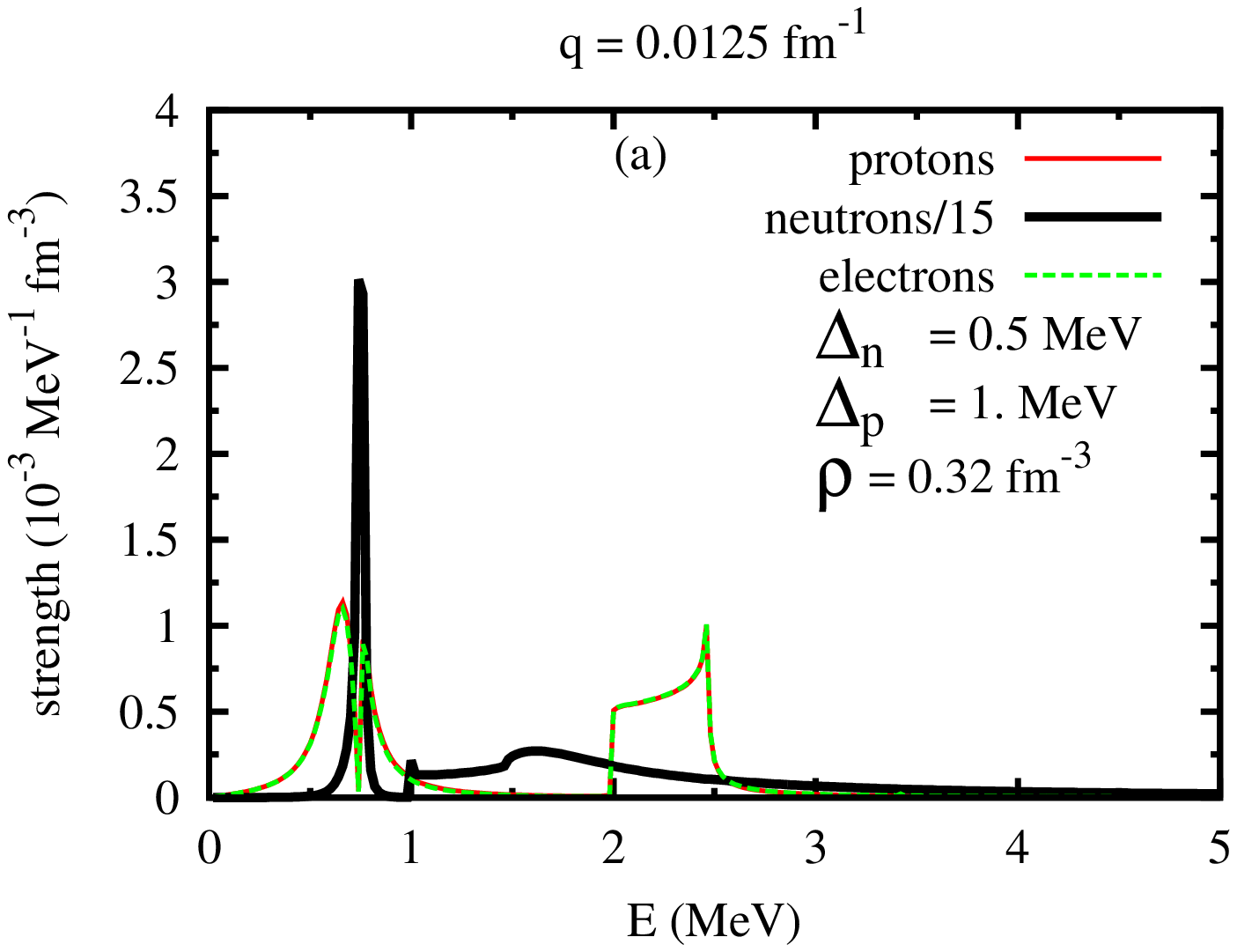}
\includegraphics[bb= 60 0 230 790,angle=0,scale=0.55]{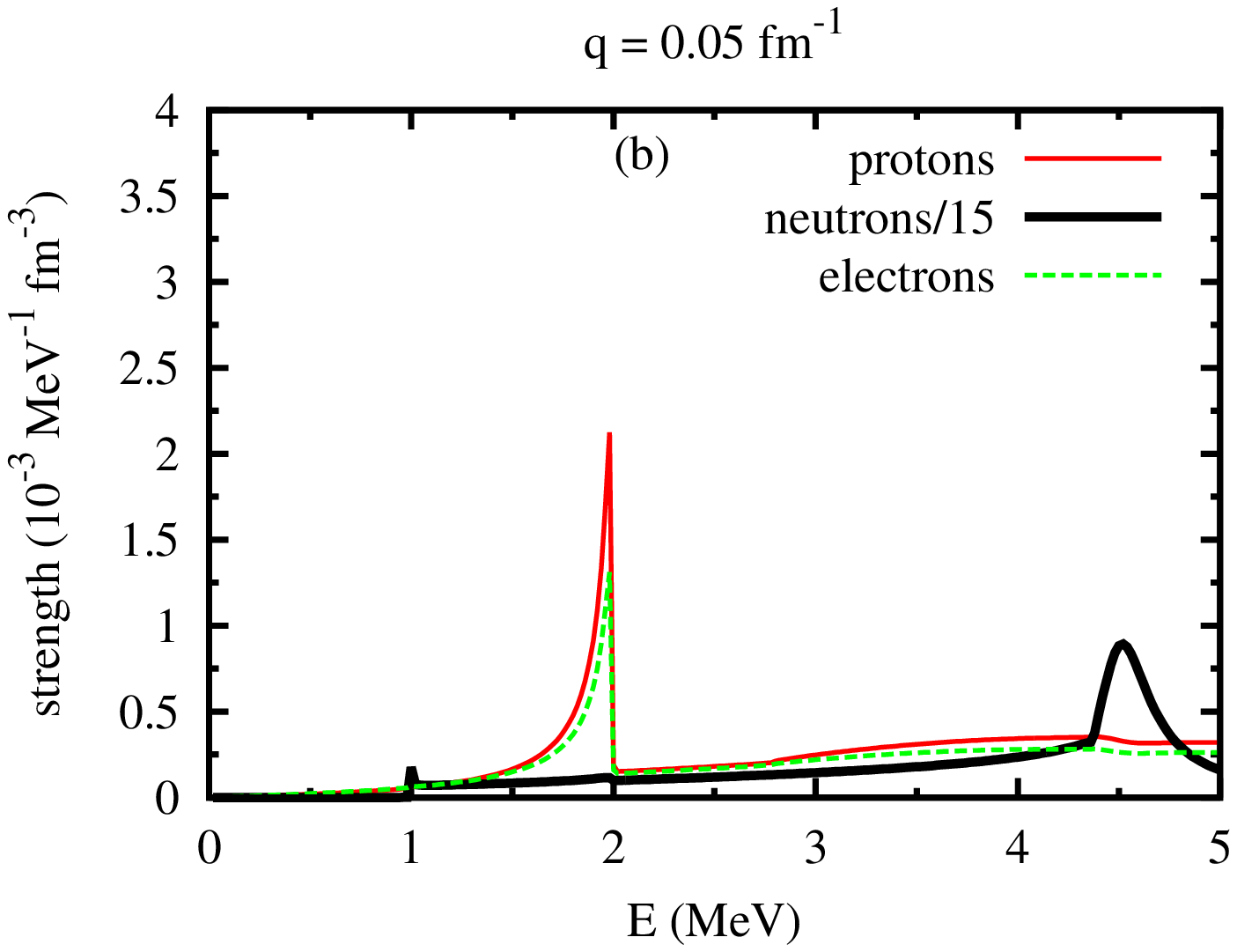}
\vskip -1 cm
\caption{(Color on line) Spectral functions of neutrons (black thick line), proton (red thin line) and electrons (green dashed line) calculated at the total baryon density $\rho = 0.32$ fm$^{-3}$ with the same pairing gaps as in Fig. \ref{fig:Fig2}, for two different momenta. In both cases the neutron-proton interaction has been included. For convenience the neutron strength function has been divided by 15.}
\label{fig:Fig5}
\end{figure}

To analyze its density dependence, the strength function has been calculated at twice saturation density. For the same pairing gap values and the neutron-proton interaction included it is reported in Fig. \ref{fig:Fig5} at two different momenta. 
The interaction is again the microscopic one, reported in Table II of ref. \cite{paper1}. At the lower momentum, panel (a), one notices the usual phonon peak, with a substantial damping, and the fragmented proton strength function. The electron strength function follows extremely closely the proton one. 
%At higher momentum the phonon peak disappears, while the neutron and proton strength functions looks completely independent. The proton strength function is characterized by a broad phonon peak, just below 2$\Delta_p$, while the neutron strength function presents a broad peak at higher energy, well above 2$\Delta_n$. 
%%%%%%%%%%
%Camille
At higher momentum, panel b), the neutron and proton strength functions look completely independent. The proton strength function is characterized by a broad phonon peak, just below 2$\Delta_p$, while the neutron strength function has no phonon peak any more. Instead, the neutron strength function presents a broad peak at higher energy, well above 2$\Delta_n$.
%%%%%%%%%%
The latter can be interpreted as the sound mode of the neutron component. In fact, as it was shown in ref. \cite{paper3}, in a superfluid liquid at increasing momenta the phonon peak disappears, while the pair-breaking mode develops in the zero-sound mode of the system, which can be strongly damped because the particle-hole channel is coupled to the pair-breaking process. 
\par
\begin{figure}
\vskip -8 cm
\includegraphics[bb= 320 0 470 790,angle=0,scale=0.55]{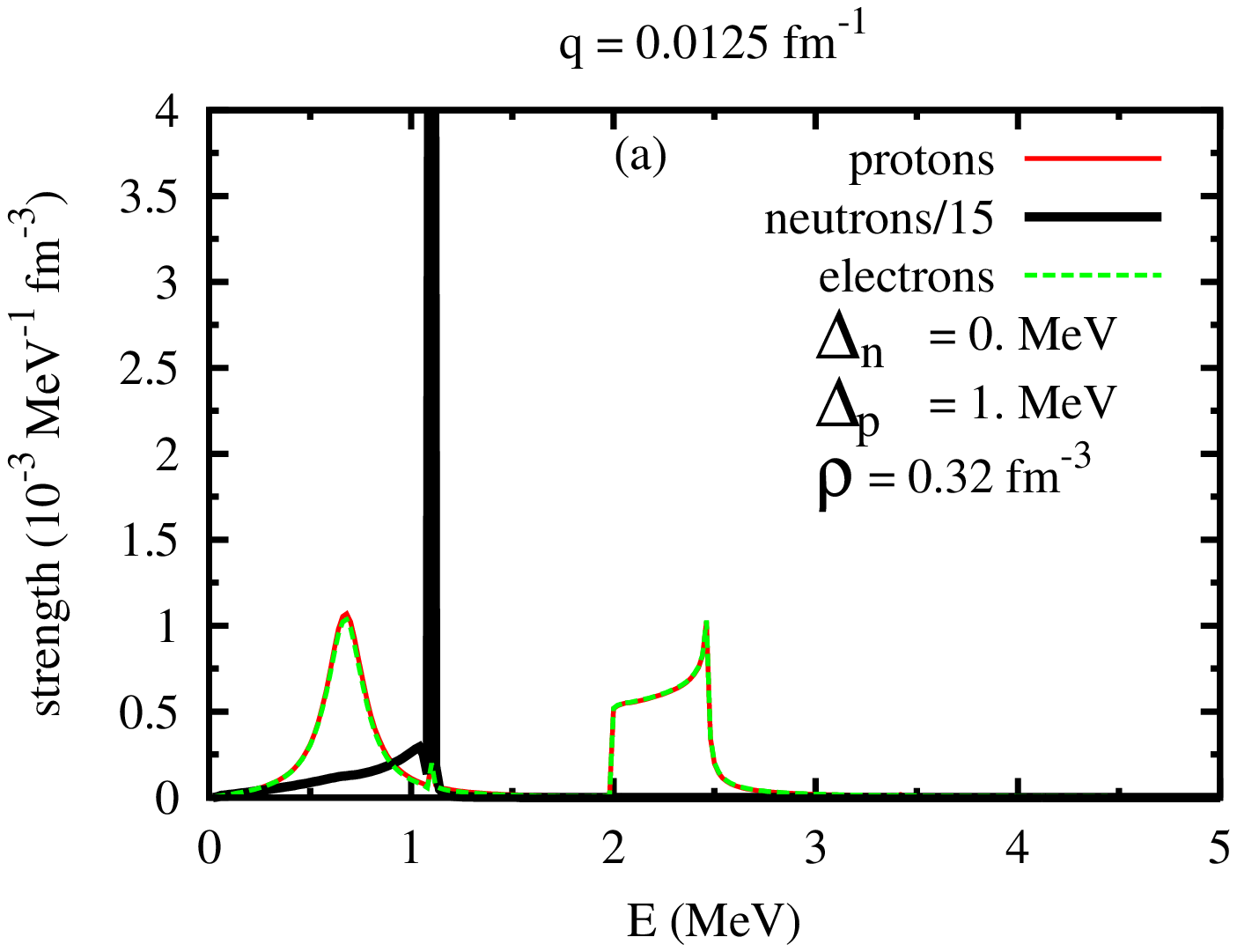}
\includegraphics[bb= 60 0 230 790,angle=0,scale=0.55]{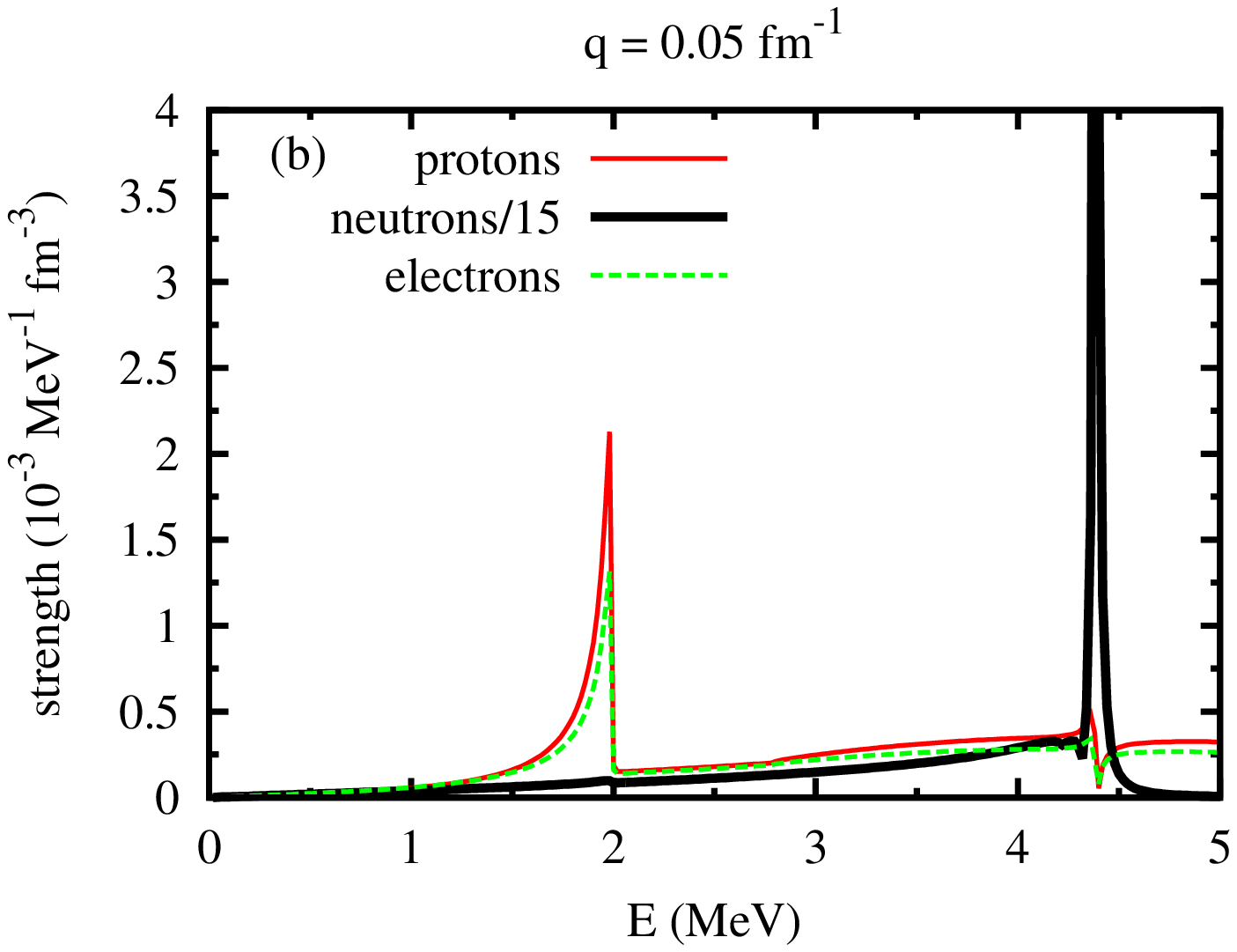}
\vskip -1 cm
\caption{(Color on line) Spectral functions of neutrons (black thick line), proton (red thin line) and electrons (green dashed line) calculated at the total baryon density $\rho = 0.32$ fm$^{-3}$ of the neutron star matter 
%with as in Fig. \ref{fig:Fig5}, 
with proton gap as in \ref{fig:Fig2}, 
but taking the neutron pairing gap as vanishing. In both cases the neutron-proton interaction has been included. For convenience the neutron strength function has been divided by 15.}
\label{fig:Fig6}
\end{figure}

To see this more clearly we have reported in Fig. \ref{fig:Fig6} the strength functions under the same physical conditions
but assuming the neutron gap vanishing small, i.e. a normal neutron component. One can see that now the broad neutron peak at the higher momentum has become a relatively narrow peak, which indicates that the apparent damping of the neutron peak in Fig. \ref{fig:Fig5}b is due to the presence of neutron pairing. 
It has to be stressed that the calculated microscopic interaction in the neutron particle-hole channel is repulsive at this density, while it is attractive at saturation density. This means that the zero-sound mode in Fig. \ref{fig:Fig6}a is shifted outside the (normal) neutron particle-hole continuum, and Landau damping is not present. The width of the mode is therefore entirely due to the coupling with the protons. This can be clearly seen in panel (a), where the peak just above the neutron particle-hole continuum is apparent. 
\par
\section{Summary and Conclusions.\label{sec:conc}}
In the homogeneous core of a neutron star the dense matter is expected to be composed mainly by neutrons, protons and electrons. The collective elementary excitations of the medium are determined in general by the coupling among the three components and the corresponding spectral function has three components. According to the different regions of the core the neutrons and the protons can be in the normal state or in the superfluid state. Extending a previous work \cite{paper4}, where only the proton component was assumed to be superfluid, we have considered the case where both neutrons and protons are superfluid, with particular emphasis on the effects of the coupling between the two components. Besides the Coulomb interaction between protons and electrons and the pairing correlation, we included the nuclear interaction among the nucleons. The effective nuclear forces were derived from Bruckner-Hartree-Fock calculations that include three-body forces and correctly reproduce the nuclear matter saturation point. If the neutron-proton interaction is neglected, both neutron and proton strength functions at low momenta are characterized by the presence of a sharp phonon excitation below twice the pairing gap and a broad pair-breaking mode above it. At higher momenta the phonon peak disappears and the pair-breaking mode merges in the zero-sound mode of the system. With the introduction of the neutron-proton interaction the phonon becomes a collective neutron-proton excitation and it is damped, i.e. the phonon peak has a substantial width. The reason of the width is in the position of the mode in the energy-momentum plane, which lies inside the electron particle-hole continuum. This introduces a Landau damping, which in turn is acting on all three components through the nuclear neutron-proton interaction and the proton-electron Coulomb interaction. All that appears clearly from the calculated spectral functions. The presence of damping can affect all physical processes 
that involve the phonon degrees of freedom inside neutron-star core.
The neutron pair-breaking mode are only marginally affected by the neutron-proton interaction. However at higher momentum the neutron zero-sound mode is strongly damped by the coupling between the particle-hole continuum and the pair-breaking processes.\par  
The neutron component in the core could be superfluid in the $^{3}$P$_2$ channel, especially at high density, and an overlap with proton superfluidity is possible. This depends of course on the extension of the proton pairing in the core. For pure neutron matter the elementary excitations have been studied in the RPA framework in ref. \cite{3P2}. It would be interesting to include the proton component and its coupling. This study is left to a future work.\par 
Finally we remark that the effective nuclear interaction among nucleons can be derived also from Skyrme effective forces. 
%However this implies an extrapolation of these forces to high density and high asymmetry, where they are not tested. In ref. \cite{paper4} it has been shown that 
%the Skyrme forces differ among each other and with the force used here. 
%We believe that the microscopic forces used in this work are more reliable, at least for qualitative conclusions. 
%%%%%%%%%
%Camille
In ref. \cite{paper4} it has been shown that the quantitative features of the spectral functions are model-dependent, with differences appearing among Skyrme forces and between these and BHF results. However, the qualitative conclusions are not affected.
It is recognized that microscopic calculations are more reliable to extrapolate the high density and asymmetry behavior of nuclear interactions, and actually, modern Skyrme forces use microscopic equations of state as constraints away from nuclear data. For these reasons, we have chosen the microscopic forces as a scheme of reference for the present work.
%%%%%%%%%
  
%\section{Appendix}
%{\bf Appendix}\vskip 0.3 cm
\appendix
\section{}
\label{ap:detail}
\par
%\numberwithin{equation}{A}
In this appendix we write down the explicit expressions of the system of equations (\ref{eq:RPA}) in the main text for the response functions $ \Pi_{ij} $. In extended form the system can be written as follows
\beq
\left(\begin{array}{ccccc}
1-X^{p}_{+}U_p & -2X_{GF}^{-}v_{pp} & 2X_{GF}^{-}v_c & -2X_{GF}^{-}v_{pn} &     0   \\[0.25cm]
X_{GF}^{p}U_p & 1-2X^{ph}_{p}v_{pp}  & 2X^{ph}_{p}v_c  & -2X^{ph}_{p}v_{pn} &      0  \\[0.25cm]
      0            &  2X^{e}v_c & 1-2X^{e}v_c &  0  &   0 \\[0.25cm]
      0            &  -2X^{ph}_{n}v_{np} &     0      & 1-2X^{ph}_{n}v_{nn} &  X_{GF}^n U_n   \\[0.25cm]
      0            & -2_{GF}^nv_{np}    &     0      & -2X_{GF}^nv_{nn}    & 1-2X^p_{+}U_n   \\[0.25cm]           
\end{array}\right) %\nonumber
\left(\begin{array}{c}
\Pi^{ (p)}_S\phantom{ph} \\[0.25cm]
\Pi^{(ph,p)}_S \\[0.25cm]
\Pi^{(ph,e)}_S\\[0.25cm]
\Pi^{(ph,n)}_S\\[0.25cm]
\Pi^{ (n)}_S\phantom{ph}
\end{array}\right)
= \left(\begin{array}{c}
\Pi^{(p)}_{0,S}\phantom{ph} \\[0.25cm]
\Pi^{(ph,p)}_{0,S} \\[0.25cm]
\Pi^{(ph,e)}_{0,S} \\[0.25cm]
\Pi^{(ph,n)}_{0,S} \\[0.25cm]
\Pi^{(n)}_{0,S}\phantom{ph} 
\end{array}\right)
\label{eq:RPA2}
\eeq
\noindent
Here we have introduced the notation:
\be
\label{eq:Xpp}
X_{+}&=&\frac{1}{2}\left[X_{GG}(q)+X_{GG}(-q)\right] + X_{FF}(q) \\
\label{eq:Xph}
X^{ph}&=&X_{GG}^{ph}(q) - X_{FF}(q) \\
\label{eq:XGF}
X_{GF}^{-}&=&X_{GF}(q)- X_{GF}(-q) 
%\\
%\label{eq:XGGC}
%X_{GGC}&=&\frac{1}{2}\left[X_{GG}^{pp}(-q)-X_{GG}^{pp}(q)\right]
\ee
valid for both protons and neutrons (superscript $p,n$ in Eq. (\ref{eq:RPA2})). The different terms are the following four-dimensional integrals:
\be
\label{eq:Xinit}
X_{GG}^{ph}(q)&=&\frac{1}{i}\int\frac{dk}{(2\pi)^4}G(k)G(k+q)\;\;;\;\;
X_{GG}^{ph}(-q)=X_{GG}^{ph}(q) \\
X_{GG}(q)&=&\frac{1}{i}\int\frac{dk}{(2\pi)^4}G(k)G(-k+q) \\
X_{GG}(-q)&=&\frac{1}{i}\int\frac{dk}{(2\pi)^4}G(k)G(-k-q) \\
X_{GF}(q)&=&\frac{1}{i}\int\frac{dk}{(2\pi)^4}G(k)F(k+q) \\
X_{GF}(-q)&=&\frac{1}{i}\int\frac{dk}{(2\pi)^4}G(k)F(k-q) \\
X_{FF}(q)&=&\frac{1}{i}\int\frac{dk}{(2\pi)^4}F(k)F(k+q)\;\;;\;\; X_{FF}(-q)=X_{FF}(q)
\label{eq:Xfin}
\ee
\noindent
In the last equations $ G $ and $ F $ are the normal and anomalous single particle Green's functions for protons and neutrons. 
 Finally $ X^e $ is the relativistic electron Lindhard function \cite{Jan} and $v_c$ is the Coulomb interaction with the positive sign. The proton-proton interaction $v_{pp}$ is assumed to include it.\par
The response functions appearing in Eq. (\ref{eq:RPA2}) are labeled according to the operator that appears on the left of the
time-ordered product of Eq. (\ref{eq:Pi}), according to the list (\ref{eq:conf}). 
% Camille
%So the label $ pp $ stands for the configuration 
%$ (a^\dag(p) a^\dag(p) \,-\, a(p) a(p)) | \, \Psi_0 \, >$, the label $ ph $ stands for $ a^\dag(p) a(p) | \, \Psi_0\, > $ or
%$ a^\dag(n) a(n) | \, \Psi_0\, > $ and finally $ (ph,e) $ for the electron particle-hole configuration. 
So the labels $ (p),(n) $ stand for the configuration 
$ (a^\dag(p) a^\dag(p) \,-\, a(p) a(p)) | \, \Psi_0 \, >$, and  $ (a^\dag(n) a^\dag(n) \,-\, a(n) a(n)) | \, \Psi_0 \, >$,
respectively, while the labels $ (ph,p) $, $ (ph,n) $ and $ (ph,e) $ 
stand for the particle-hole configurations 
$ a^\dag(p) a(p) | \, \Psi_0\, > $,
$ a^\dag(n) a(n) | \, \Psi_0\, > $ and 
$ a^\dag(e) a(e) | \, \Psi_0\, > $.
%%%%%%%%%%%
The same labeling is used for the free response functions $ \Pi_0 $ appearing on the right hand side. 
Let us stress that the quantity $\Pi_0$ and $\Pi$ are the free and the full response function respectively, see Eq. (\ref{eq:Pi}),
and therefore they are 5x5 matrices. The index $j$ on the right side of both response functions in Eq. (\ref{eq:RPA}) is not indicated here for simplicity of notation, since it is mainly a dummy index and it can indicate any one of the above mentioned configurations.
For each choice of the configuration $j$ on the right side
%of the time-ordered product of Eq. (\ref{eq:Pi}) 
the free response functions will change properly and the corresponding correlated response functions can be calculated. In this way the whole 5x5 matrix $\Pi$ can be obtained. The subscript $ S $ in the response functions specifies that they are calculated for the spin zero (scalar)  case.  \par 
The analytic expressions for the integrals of Eqs. (\ref{eq:Xinit}-\ref{eq:Xfin}) can be found in ref. \cite{paper3}. 
%%%%%%%%%%%%%%%%%%%%%%%%%%%%%%%%%%%%%%%%%%%%%%%%%%%%%%%%%%%%%%%%%%%%%%%%%%%%%%%
%%%%%%%%%%%%%%%%%%%%%%%%%%%%%%%%%%%%%%%%%%%%%%%%%%%%%%%%%%%%%%%%%%%%%%%%%%%%%%%
%%%%%%%%%%%%%%%%%%%%%%%%%%%%%%%%%%%%%%%%%%%%%%%%%%%%%%%%%%%%%%%%%%%%%%%%%%%%%%%

\end{document}